\documentclass[letterpaper,10pt,conference]{ieeeconf}

\IEEEoverridecommandlockouts

\usepackage{graphicx} 
\usepackage{amsmath,amssymb,amsthm}
\usepackage{tikz}
\usetikzlibrary{
    arrows.meta,
    calc,
    positioning,
    shapes.geometric,
    backgrounds
}

\definecolor{leaderblue}{RGB}{30,105,190}
\definecolor{escortorange}{RGB}{230,125,25}
\definecolor{pursuerred}{RGB}{185,45,45}
\definecolor{openinggreen}{RGB}{70,150,85}
\definecolor{terrain}{RGB}{224,216,195}

\usepackage{xcolor}

\newtheorem{theorem}{Theorem}
\newtheorem{lemma}{Lemma}
\newtheorem{proposition}{Proposition}

\newtheorem{remark}{Remark}
\title{\LARGE\bf Minimum-Time Trajectory for Escort-Assisted Evasion from a Sensing-Limited Pursuer}
\author{%
Al Faysal, Alexander Von Moll, Artur Wolek, Satyanarayana G. Manyam, Thomas Chapman, Isaac Weintraub,\\  David Casbeer, and Dipankar Maity
\thanks{%
A. Faysal, A. Wolek, and D. Maity are with the University of North Carolina at Charlotte, Charlotte, NC, 28223, USA.%
}
\thanks{%
A. Von Moll, T. Chapman, I. Weintraub, and D. Casbeer are with Air Force Research Laboratory, Wright-Patterson AFB, OH 45433, USA.%
}
\thanks{%
S. G. Manyam is with DCS Corporation.%
}
\thanks{%
DISTRIBUTION STATEMENT A. Approved for public release. Distribution is unlimited. AFRL-2026-1762; Cleared 6 Oct 2026.%
}
}

\newcommand{\xg}{\mathbf{x}_g}

\newcommand{\xlo}{\mathbf{x}_{L0}}
\newcommand{\xeo}{\mathbf{x}_{E0}}
\newcommand{\xpo}{\mathbf{x}_{P0}}

\newcommand{\tfua}{t_f^{\rm unassisted}}

\begin{document}

\maketitle

\begin{abstract}
This paper studies a minimum-time path-planning problem in which an escort assists a leader by triggering and redirecting a sensing-limited pursuer. For the post-engagement case with equal leader, escort, and pursuer speeds, we show that the minimum-time leader trajectory has a three-phase structure: an initial straight segment, a moving-boundary segment, and a final straight segment to the goal. The boundary-following kinematics and the
corresponding entry and exit conditions are derived and used to construct the leader trajectory for a fixed escort engagement direction. The engagement direction is then optimized over the feasible set to obtain the minimum-time solution. Numerical results illustrate the resulting trajectory and the benefits of escort assistance.
\end{abstract}

\section{Introduction}
\label{sec:introduction}
Autonomous vehicles are increasingly required to operate in environments containing other moving agents, obstacles, or threats. In many such missions, an agent must reach a desired destination while maintaining safety with respect to these entities. The use of multiple autonomous agents can provide additional flexibility in such environments. In particular, one agent may assist another by sensing, intercepting, distracting, or redirecting a moving threat. Such cooperative behaviors arise in applications involving unmanned aerial/ ground / surface vehicles, including surveillance, search and rescue, infrastructure protection, environmental monitoring, and maritime missions \cite{Chung2018,Yan2023}.

These scenarios are naturally related to pursuit--evasion and reach--avoid problems, in which one or more agents seek to reach a target while avoiding interception or capture by opposing agents. Such problems have been studied extensively using both numerical and analytical approaches \cite{Isaacs1965,Weintraub2020}. Hamilton--Jacobi--Isaacs methods provide a general framework for characterizing reachable and winning sets \cite{Mitchell2005}. However, the computational complexity of these methods grows rapidly with the dimension of the joint state space, particularly in multi-agent problems \cite{Gao2024}. This has motivated the development of geometric and analytical approaches for problem classes in which the structure of the optimal trajectories can be characterized explicitly \cite{ChenZhouTomlin2017,GarciaCasbeerPachter2020}.

Limited sensing by the opposing agent can also be exploited strategically \cite{Gerkey2006,Durham2010}. In such settings, the pursuer may not react to another agent until that agent enters its sensing region. Consequently, the pursuer's information and subsequent motion may depend on the time and location at which sensing occurs. This leads to sensing-limited pursuit--evasion problems, where the sensing geometry becomes an important part of the optimal strategy \cite{Shishika2021,MaityEtAl2024}.

Limited sensing can also be exploited cooperatively. A supporting autonomous agent may deliberately trigger a sensing-limited opponent and influence its subsequent motion, thereby enabling a more favorable route for its teammate. Motivated by this idea, we consider a leader--escort--pursuer system in which the leader seeks to reach a prescribed goal while maintaining  safety with respect to the pursuer, and the escort assists the leader by triggering and redirecting the pursuer.

The main contributions of this work are as follows. First, we characterize the minimum-time trajectory structure for the case in which the escort is initially located on the pursuer's sensing boundary, so that engagement occurs immediately. We derive the corresponding straight–boundary–straight trajectory and the conditions for entering and leaving the safety boundary. Second, we characterize the leader's reachable set and use it together with the immediate-engagement solution to extend the analysis to the delayed escort engagement. This yields a constructive procedure for computing the minimum-time trajectory and the associated escort engagement.

The remainder of the paper is organized as follows. Section~II formulates the escort-assisted minimum-time problem. Section~III characterizes the optimal trajectory structure for immediate engagement. Section~IV develops the trajectory construction and the associated value function. Section~V develops the characterization of the reachable set and  extends the construction of the trajectory to delayed engagement. Section~VI presents the numerical results and Section~VII concludes the paper.

\section{Problem Formulation}
\label{sec:problem}
We consider a two-dimensional\footnote{The analysis can be extended analogously to three and higher dimensions.} setting with three agents with three different roles: a leader $L$, an escort $E$, and a pursuer $P$. The leader and the escort work as a team, and the presence of the pursuer is a threat to them. The leader starts from $\mathbf{x}_{L0}\in\mathbb{R}^2$ and seeks to reach a fixed goal $\mathbf{x}_g\in\mathbb{R}^2$ in minimum time while maintaining a prescribed safety separation from the pursuer. The pursuer is initially stationary and has a circular sensing region of radius $r$. Once an agent enters the sensing region, the pursuer is triggered and subsequently follows a pure-pursuit strategy toward that agent. The escort exploits this behavior by intentionally entering the sensing region and causing the pursuer to move; we refer to this event as \emph{engagement}. By selecting the engagement location, the escort redirects the subsequent motion of the pursuer and can thereby create a more favorable path for the leader to reach the goal. We assume that the goal location $\xg$ is unknown to the pursuer, and therefore it cannot simply  guard the goal.

Let $\mathbf{x}_i(t)\in\mathbb{R}^2$, $i\in\{L,E,P\}$, denote the position of each agent. For any angle $\alpha$, measured counterclockwise from the positive $x$-axis, define the unit direction
vector 
\[
    \mathbf{u}_{\alpha}
    \triangleq
    \begin{bmatrix}
        \cos\alpha\\
        \sin\alpha
    \end{bmatrix}.
\]
When in motion, all agents obey first-order constant-speed kinematics
\begin{equation}
    \dot{\mathbf{x}}_i(t)
    = v_i\mathbf{u}_{\psi_i(t)},
    \qquad i\in\{L,E,P\},
    \label{eq:agent_dynamics}
\end{equation}
where $v_i>0$ is the constant speed of the agent $i$ and $\psi_i(t)$ is its heading angle. In this work, we consider the equal-speed case
\begin{equation}
    v_L=v_E=v_P=v,
    \label{eq:equal_speed}
\end{equation}
where $v>0$ denotes the common speed. To impose a safety separation between the leader and the pursuer, we introduce a margin $\epsilon>0$ beyond the pursuer's sensing radius and define
\begin{equation}
    R \triangleq r+\epsilon .
    \label{eq:safety_radius}
\end{equation}
The leader is required to satisfy the following.
\begin{equation}
    \|\mathbf{x}_L(t)-\mathbf{x}_P(t)\|\geq R,
    \qquad 0\leq t\leq t_f,
    \label{eq:safety_constraint}
\end{equation}
where $t_f$ denotes its arrival time at the goal.

We restrict our attention to problem instances in which the leader is strictly outside the safety boundary at both the initial and terminal times,
\begin{equation}
    \|\mathbf{x}_{L0}-\mathbf{x}_{P0}\|>R,
    \qquad
    \|\mathbf{x}_g-\mathbf{x}_P(t_f)\|>R.
    \label{eq:endpoint_safety}
\end{equation}
Thus, the safety constraint is inactive at the endpoints, while it may become active during the leader's motion. Initial or terminal configurations on the safety boundary are not considered in this work.
\begin{figure}[t]
    \centering

\begin{tikzpicture}[
    x=0.68cm,
    y=0.68cm,
    >=Latex,
    line cap=round,
    line join=round,
    font=\scriptsize,
    labelbox/.style={
        fill=white,
        fill opacity=0.90,
        text opacity=1,
        inner sep=1.2pt,
        rounded corners=0.5pt
    }
]

\coordinate (P)  at (0,0);          
\coordinate (L0) at (-4.65,-2.35);  
\coordinate (G)  at (4.45,2.55);    

\coordinate (E) at ($(P)+(12:1.15)$);

\fill[
    green!35,
    opacity=0.15
]
    (1.05,-0.35)
    .. controls (1.65,0.10) and (2.05,0.80) ..
    (2.75,1.30)
    .. controls (3.40,1.78) and (4.00,2.15) ..
    (4.62,2.42)
    --
    (4.82,2.87)
    .. controls (3.90,2.78) and (3.00,2.40) ..
    (2.25,1.90)
    .. controls (1.58,1.42) and (1.12,0.62) ..
    (0.78,0.20)
    -- cycle;

\draw[
    green!45!black,
    dashed,
    line width=0.70pt
]
    (0.78,0.20)
    .. controls (1.60,1.35) and (3.05,2.24) ..
    (4.82,2.87);


\fill[
    red!35,
    opacity=0.22
]
    (P) circle (1.15);

\draw[
    red!75!black,
    very thick
]
    (P) circle (1.15);

\draw[
    red!70!black,
    dashed,
    line width=1pt
]
    (P) circle (1.65);

\draw[
    red!70!black,
    -{Latex[length=1.5mm]},
    line width=0.70pt
]
    (P) -- (E);

\coordinate (Rpoint) at ($(P)+(145:1.65)$);

\draw[
    red!65!black,
    -{Latex[length=1.4mm]},
    line width=0.65pt
]
    (P) -- (Rpoint);

\node[
    star,
    star points=8,
    star point ratio=0.55,
    minimum size=1mm,
    inner sep=0pt,
    fill=red!80!black,
    draw=black,
    line width=0.45pt
] at (P) {};

\node[
    regular polygon,
    regular polygon sides=3,
    minimum size=5.4mm,
    inner sep=0pt,
    fill=orange!85,
    draw=black,
    line width=0.45pt,
    rotate=-78
] at (E) {};

\fill[black] (E) circle (1.0pt);

\draw[
    red!75!black,
    very thick,
    -{Latex[length=2.2mm]}
]
    ($(E)+(0.10,0.02)$)
    -- ++(12:1.72);

\node[
    regular polygon,
    regular polygon sides=3,
    minimum size=6mm,
    inner sep=0pt,
    fill=blue!70,
    draw=black,
    line width=0.45pt,
    rotate=-68
] at (L0) {};

\draw[
    gray!70,
    dashed,
    line width=0.80pt
]
    ($(L0)+(0.18,0.13)$)
    --
    ($(G)+(-0.12,-0.07)$);

\draw[
    blue!80!black,
    very thick,
    -{Latex[length=2.2mm]}
]
    (L0)
    .. controls (-3.55,-2.72) and (-2.45,-2.60) ..
    (-1.55,-1.82)
    .. controls (-0.90,-1.24) and (-0.30,-1.18) ..
    (0.35,-1.05)
    .. controls (1.10,-0.85) and (1.55,-0.05) ..
    (1.72,0.65)
    .. controls (1.95,1.22) and (2.85,1.75) ..
    (3.45,2.05)
    .. controls (3.85,2.25) and (4.15,2.40) ..
    (G);

\draw[
    blue!80!black,
    -{Latex[length=1.5mm]},
    line width=0.60pt
]
    (-2.82,-2.54) -- (-2.38,-2.45);

\draw[
    blue!80!black,
    -{Latex[length=1.5mm]},
    line width=0.60pt
]
    (3.00,1.83) -- (3.38,2.05);

\node[
    star,
    star points=5,
    star point ratio=0.45,
    minimum size=1.8mm,
    inner sep=0pt,
    fill=yellow!90!orange,
    draw=black,
    line width=0.45pt
] at (G) {};


\node[
    labelbox,
    anchor=west,
    text=blue!70!black
] at (-4.32,-2.34)
    {\textbf{Leader}};

\node[
    labelbox,
    anchor=west,
    text=blue!70!black
] at (-4.15,-2.70)
    {$x_{L0}$};

\node[
    labelbox
] at (0,-0.53)
    {\textbf{Pursuer}};

\node[
    labelbox
] at (0,-0.91)
    {$x_{P0}$};

\node[
    labelbox,
    text=red!65!black
] at (0.58,0.44)
    {$r$};

\node[
    labelbox,
    text=red!65!black
] at (-1.60,1.36)
    {$R=r+\epsilon$};

\node[
    labelbox,
    align=center
] at (-2.35,0.35)
    {direct path\\infeasible};

\node[
    labelbox,
    anchor=south
] at (0.80,0.69)
    {$x_{\mathrm{eng}}(\theta)$};

\node[
    labelbox,
    anchor=west
] at (1.43,0.83)
    {\textbf{Escort}};

\node[
    labelbox
] at (1.63,0.22)
    {$\mathbf{u}_{\theta}$};

\node[
    labelbox,
    text=green!40!black
] at (3.05,2.54)
    {\textbf{opening}};

\node[
    labelbox,
    anchor=west
] at (4.53,2.76)
    {\textbf{Goal}};

\node[
    labelbox,
    anchor=west
] at (4.53,2.42)
    {$x_g$};

\end{tikzpicture}

\caption{
Illustration of escort-assisted path planning. The escort engages the sensing-limited pursuer at $x_{\mathrm{eng}}(\theta)$ and redirects its subsequent motion, creating an opening for the leader to reach the goal while maintaining the safety separation $R=r+\epsilon$.
}
\label{fig:escort_concept}

\end{figure}
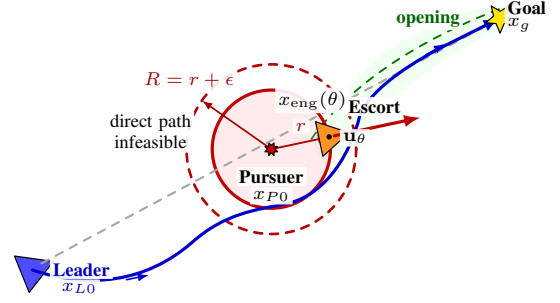

\subsection{Escort Engagement and Pursuer Motion}
Let $\mathbf{x}_{E0}\in\mathbb{R}^2$ and $\mathbf{x}_{P0}\in\mathbb{R}^2$ denote the initial positions of the escort and the pursuer, respectively. The escort selects an engagement location on the pursuer's sensing boundary, parameterized by the angle $\theta\in[0,2\pi)$ as
\begin{equation}
    \mathbf{x}_{\mathrm{eng}}(\theta)
    =
    \mathbf{x}_{P0}+r\mathbf{u}_{\theta}.
    \label{eq:engagement_point}
\end{equation}
The angle $\theta$ is therefore referred to as the \emph{engagement angle}.

For a prescribed $\theta$, let $T_{\mathrm{eng}}(\theta)$ denote the minimum time required for the escort to reach $\mathbf{x}_{\mathrm{eng}}(\theta)$ without entering the interior of the pursuer's sensing region prior to engagement. The corresponding minimum-time escort path is either a direct straight segment to $\mathbf{x}_{\mathrm{eng}}(\theta)$ or, when the direct segment intersects the sensing-region interior, a tangent segment followed by motion along the sensing boundary. Tangential contact with the sensing boundary prior to $\mathbf{x}_{\mathrm{eng}}(\theta)$ is not regarded as engagement.

Upon engagement, both the pursuer and the escort move with speed $v$ in the direction
$\mathbf{u}_{\theta}$. Therefore,
\begin{equation}
\mathbf{x}_P(t;\theta)=
\begin{cases}
    \mathbf{x}_{P0},
        & 0\leq t\leq T_{\mathrm{eng}}(\theta),\\[1mm]
    \mathbf{x}_{P0}
    +v\bigl(t-T_{\mathrm{eng}}(\theta)\bigr)\mathbf{u}_{\theta},
        & t>T_{\mathrm{eng}}(\theta).
\end{cases}
\label{eq:pursuer_motion}
\end{equation}
Thus, for a fixed engagement angle $\theta$, the leader encounters a circular safety region of radius $R$ whose center is stationary before engagement and translates with constant velocity $v\mathbf{u}_{\theta}$ after engagement.

\begin{remark}
In case the assumption of equal speed is violated and the escort has a speed smaller than the pursuer, then the escort will eventually be captured at the location $\mathbf{x}_{\rm capture} = \mathbf{x}_{P0} + \tfrac{r v_P}{v_P - v_E}\mathbf{u}_\theta$, and its assistance to the leader ends. In either case, equal or unequal speeds, the motion of the pursuer is deterministic and known to the leader, which will be utilized to find the optimal leader trajectory. We continue with the equal speed assumption for simplicity of the analysis and the exposition. \end{remark}

\subsection{Leader Minimum-Time Problem}
For a prescribed engagement angle $\theta$, the pursuer trajectory $\mathbf{x}_P(t;\theta)$ is completely determined by \eqref{eq:pursuer_motion}. The leader then chooses its heading $\psi_L(\cdot)$ and free terminal time $t_f$ to reach the goal in minimum time while satisfying the safety constraint. The resulting fixed-$\theta$ problem is
\begin{equation}
\small
\begin{aligned}
T^\star(\theta)
=\min_{\psi_L(\cdot),\,t_f}\;& t_f\\
\mathrm{s.t.}\quad
&\dot{\mathbf{x}}_L=v\mathbf{u}_{\psi_L},\\
&\mathbf{x}_L(0)=\mathbf{x}_{L0}, 
\qquad\mathbf{x}_L(t_f)=\mathbf{x}_g,\\
&\|\mathbf{x}_L(t)-\mathbf{x}_P(t;\theta)\|\ge R,
\quad 0\le t\le t_f,\\
&\|\mathbf{x}_g-\mathbf{x}_P(t_f;\theta)\|>R .
\end{aligned}
\label{eq:fixed_theta_problem}
\end{equation}
We restrict attention to engagement angles for which the corresponding minimum travel time is smaller than the unassisted travel time. 

Let $\tfua(\xg, \xlo)$ denote the minimum time to reach $\xg$ from $\xlo$ when the pursuer remains untriggered. That is, the escort's assistance is disregarded. Accordingly, we define
\begin{equation}
\Theta_{\mathrm{feas}}
\triangleq
\left\{
\theta\in[0,2\pi):
T^\star(\theta)< \tfua(\xg, \xlo)
\right\}.
\label{eq:theta_feasible}
\end{equation}

The overall minimum-time problem is therefore
\begin{equation}
T^\star
=\min_{\theta\in\Theta_{\mathrm{feas}}} T^\star(\theta),
    \qquad
    \theta^\star
    \in
    \arg \min_{\theta\in\Theta_{\mathrm{feas}}}
    T^\star(\theta).
\label{eq:outer_optimization}
\end{equation}
We first solve the leader's minimum-time trajectory problem for a fixed engagement angle $\theta$, and then optimize over the escort-selected engagement angle.
\begin{figure*}[t]
    \centering
    \includegraphics[width=\textwidth]{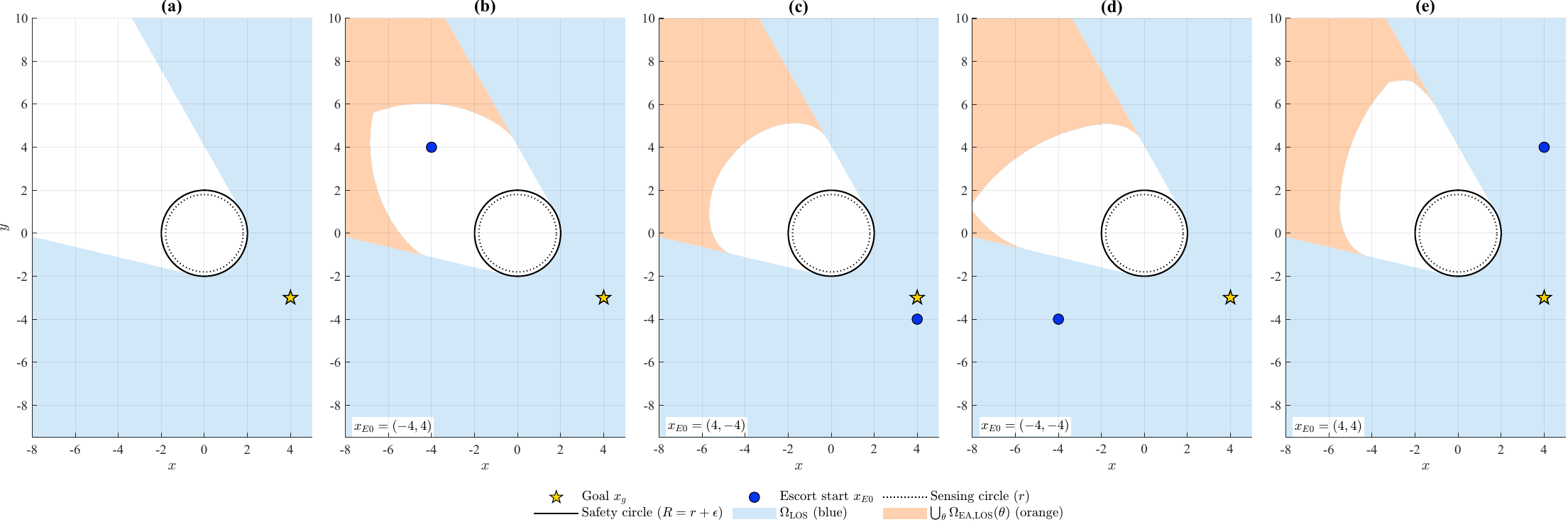}
    \caption{Straight-line safe regions for different
    escort configurations.
    (a) Direct line-of-sight safe region
    $\Omega_{\mathrm{LOS}}$ without escort assistance.
    (b)--(e) Escort-assisted straight-line safe regions
    $\bigcup_{\theta\in[0,2\pi)}
    \Omega_{\mathrm{EA,LOS}}(\theta)$
    for different escort initial positions.}
    \label{fig:Los}
\end{figure*}

\subsection{Straight-Line Feasibility}
The straight-line feasibility conditions for the leader, both without escort assistance and with engagement at a prescribed angle $\theta$, were characterized in \cite{maityminimum2026}. Following that result, define
\begin{equation}
    \Omega_{\mathrm{straight}}
    \triangleq
    \Omega_{\mathrm{LOS}}
    \cup
    \bigcup_{\theta\in[0,2\pi)}
    \Omega_{\mathrm{EA,LOS}}(\theta),
    \label{eq:straight_region}
\end{equation}
where $\Omega_{\mathrm{LOS}}$ denotes the set of initial leader positions admitting a safe direct path without escort assistance, and $\Omega_{\mathrm{EA,LOS}}(\theta)$ denotes the corresponding set when the escort engages the pursuer at angle $\theta$; see Fig.~\ref{fig:Los}. For $\mathbf{x}_{L0}\in\Omega_{\mathrm{straight}}$, the straight-line path to the goal is minimum-time \cite{maityminimum2026}. Therefore, the remainder of this paper considers
\begin{equation}
    \mathbf{x}_{L0}\notin\Omega_{\mathrm{straight}},
    \label{eq:curved_case}
\end{equation}
for which no safe straight-line trajectory to the goal exists.

\section{Optimal Trajectory Structure: Pursuer is triggered}
\label{sec:optimal_structure}
Let us start the analysis at the moment the pursuer is triggered. The escort is at a given angular location $\theta$ on the pursuer's sensing boundary, i.e., $\xeo = \xpo + r \mathbf{u}_\theta$.

The active set consists of the times at which the safety constraint is active.
\begin{equation}
   \mathcal{A}
   \triangleq
   \left\{
         t\in[0,t_f]:
         \|\mathbf{x}_L(t)-\mathbf{x}_P(t)\|=R
    \right\}.
\end{equation}
An interval over which $\|\mathbf{x}_L(t)-\mathbf{x}_P(t)\|>R$ is referred to as an \textit{unconstrained arc}.

\begin{lemma}[Straight Unconstrained Arcs]
For a given engagement angle $\theta$, every unconstrained arc of a minimum-time leader trajectory is a straight-line segment.
\end{lemma}

\begin{proof}
Define the relative state
\begin{equation}
    \mathbf{z}
    \triangleq
    \mathbf{x}_L-\mathbf{x}_P .
\end{equation}
Since the pursuer moves with constant velocity $v\mathbf{u}_\theta$, the relative equations of motion are
\begin{equation}
\dot{\mathbf{z}}
=
v\bigl(\mathbf{u}_{\psi_L}-\mathbf{u}_\theta\bigr).
\end{equation}

Let $I$ be an unconstrained arc. Since the safety constraint is inactive on $I$, the minimum-time running cost is $L=1$. Hence, the normal\footnote{The exclusion of abnormal extremals and the nonvanishing of the costate on unconstrained arcs are established in the Appendix.~\ref{app:normality}} Hamiltonian is
\begin{equation}
H_R
=
1+\mathbf{p}^{\top}
v\bigl(\mathbf{u}_{\psi_L}-\mathbf{u}_\theta\bigr), \label{eq:HR}
\end{equation}
where $\mathbf{p}$ is the costate associated with $\mathbf{z}$. Because $H_R$ is independent of $\mathbf{z}$,
\begin{equation}
\dot{\mathbf{p}}
=
-\frac{\partial H_R}{\partial \mathbf{z}}
=
\mathbf{0}.
\label{eq:costate_dot}
\end{equation}
Hence, $\mathbf{p}$ is constant on $I$.

The Pontryagin minimum principle requires $\mathbf{u}_{\psi_L}$ to minimize $H_R$~\cite{BrysonHo1975}. Since $\mathbf{u}_\theta$ is constant and $\mathbf{p}\neq\mathbf{0}$,
\begin{equation}
\mathbf{u}_{\psi_L}^{\star}
=
-\frac{\mathbf{p}}{\|\mathbf{p}\|}.
\label{unconstrained_u}
\end{equation}
Thus, $\mathbf{u}_{\psi_L}^{\star}$ is constant on $I$  and the leader follows a straight-line segment on~$I$.
\end{proof}

To exclude chattering at the safety boundary, we restrict attention to optimal trajectories with finitely many boundary junctions. Any interval of positive duration contained in the active set $\mathcal A$ is referred to as a boundary arc.

\begin{lemma}[Existence of a Boundary Arc]
The minimum-time trajectory contains a boundary arc of positive duration; that is, there exist times $t_e<t_x$ such that
\begin{equation}
\|\mathbf{x}_L(t)-\mathbf{x}_P(t)\|=R,
\qquad t\in[t_e,t_x].
\end{equation}
\end{lemma}

\begin{proof}
Suppose, by contradiction, that the active set $\mathcal{A}$ contains no interval of non-zero duration. If $\mathcal{A}=\varnothing$, the entire trajectory is unconstrained. By Lemma~1, the leader follows a single straight-line segment from $\mathbf{x}_{L0}$ to $\mathbf{x}_g$, contradicting the standing assumption that the direct straight-line trajectory is infeasible due to \eqref{eq:curved_case}.

It remains to consider the case in which $\mathcal{A}$ consists of isolated contacts, i.e., individual time instants at which the safety constraint is active but no boundary arc occurs. Let $t_c$ be one such contact and define
\begin{equation}
\mathbf{z}_c\triangleq
\mathbf{x}_L(t_c)-\mathbf{x}_P(t_c)
=
R\mathbf{n}_c,
\end{equation}
where $\mathbf{n}_c$ is the outward unit normal to the safety boundary, pointing away from the pursuer. Introduce the state constraint 
\begin{equation}
g(\mathbf{z})=R^2-\|\mathbf{z}\|^2\leq 0.
\end{equation}
As shown in the Appendix, abnormal extremals are excluded. The state-constraint corner conditions ~\cite{VinterZheng1998},~\cite{vinter2000optimal} at $t_c$ therefore give
\begin{equation}
\mathbf{p}(t_c^-)
=
\mathbf{p}(t_c^+)
+
\mu_c\nabla g(\mathbf{z}_c)
=
\mathbf{p}(t_c^+)
-
2\mu_c R\mathbf{n}_c,
\qquad
\mu_c\geq 0,
\label{eq:isolated_costate_jump} 
\end{equation}
Here, $p(t_c^-)$ and $p(t_c^+)$ denote the left- and right-hand limits of the costate at the contact time $t_c$ and $\mu_c$ is the multiplier associated with the isolated contact.
Since $g$ has no explicit time dependence in the relative coordinates
\begin{equation}
H_R(t_c^-)=H_R(t_c^+),
\label{eq:isolated_H_continuity}
\end{equation}

Recall from \eqref{unconstrained_u} that $\mathbf{u}^*(t) = -\frac{\mathbf{p}(t)}{\|\mathbf{p}(t)\|}$.  Let us define
\[
\mathbf{u}^{-}
\triangleq
\mathbf{u}^*_{\psi_L(t_c^-)} = -\frac{\mathbf{p}(t_c^-)}{\|\mathbf{p}(t_c^-)\|},
\quad
\mathbf{u}^{+}
\triangleq
\mathbf{u}^*_{\psi_L(t_c^+)} = -\frac{\mathbf{p}(t_c^+)}{\|\mathbf{p}(t_c^+)\|}.
\] 
For notational convenience, let us further define
\[
\kappa\triangleq 2\mu_cR,
\qquad
a\triangleq\mathbf{n}_c^\top\mathbf{u}_\theta,
\qquad
b\triangleq\mathbf{n}_c^\top\mathbf{u}^{+}, 
\]
\[
\rho^-\triangleq {\|\mathbf{p}(t_c^-)\|}, \qquad \rho^+\triangleq {\|\mathbf{p}(t_c^+)\|}.
\]
Thus, equation~\eqref{eq:isolated_costate_jump} becomes
\begin{equation}
\rho^{-}\mathbf{u}^{-}
=
\rho^{+}\mathbf{u}^{+}
+
\kappa\mathbf{n}_c.
\label{eq:isolated_vector_relation}
\end{equation}

Using \eqref{eq:HR} and \eqref{eq:isolated_H_continuity}, we obtain
\begin{equation}
\rho^{-}
\left(
1-(\mathbf{u}^{-})^\top\mathbf{u}_\theta
\right)
=
\rho^{+}
\left(
1-(\mathbf{u}^{+})^\top\mathbf{u}_\theta
\right).
\label{eq:isolated_H_relation}
\end{equation}
Taking the inner product of \eqref{eq:isolated_vector_relation} with $\mathbf{u}_\theta$ and combining the result with \eqref{eq:isolated_H_relation} gives
\begin{equation}
\rho^{-}=\rho^{+}+\kappa a.
\label{eq:rho_relation}
\end{equation}
Taking the squared norm of \eqref{eq:isolated_vector_relation} and using \eqref{eq:rho_relation} yields
\begin{equation}
\kappa
\left[
2\rho^{+}(b-a)
+
\kappa(1-a^2)
\right]
=0.
\label{eq:isolated_multiplier_relation}
\end{equation}

Because $t_c$ is an isolated contact, feasibility immediately after
$t_c$ requires
\begin{equation} \label{eq:g_dot}
    \frac{\mathrm{d}}{\mathrm{d}t}g(\mathbf{z}(t_c^+))\leq 0.
\end{equation}
Since $\dot g
=
-2\mathbf{z}^{\top}\dot{\mathbf{z}}$, we have
\begin{equation} \label{eq:g_dot_expanded}
\dot g(t_c^+) 
=
-2Rv\,\mathbf{n}_c^\top
(\mathbf{u}^+-\mathbf{u}_\theta)
=
-2Rv(b-a).
\end{equation}
Combining \eqref{eq:g_dot} and \eqref{eq:g_dot_expanded} yields
\begin{equation}
b-a
=
\mathbf{n}_c^\top
(\mathbf{u}^{+}-\mathbf{u}_\theta)
\geq 0.
\label{eq:post_contact_feasibility}
\end{equation}
Moreover, $|a|\leq 1$. Hence both terms inside the brackets in \eqref{eq:isolated_multiplier_relation} are nonnegative.

If $\kappa>0$, then \eqref{eq:isolated_multiplier_relation} requires
\[
b=a,
\qquad
|a|=1.
\]
Since $\mathbf{u}^{+}$, $\mathbf{u}_\theta$, and $\mathbf{n}_c$ are unit vectors, these equalities imply
\[
\mathbf{u}^{+}=\mathbf{u}_\theta.
\]
Consequently,
\[
\dot{\mathbf{z}}(t_c^+)
=
v(\mathbf{u}^{+}-\mathbf{u}_\theta)
=
\mathbf{0}
\]
on the unconstrained arc immediately following $t_c$, so the leader would remain on the safety boundary, contradicting the assumption that $t_c$ is an isolated contact. Therefore,
\[
\kappa=0,
\qquad\text{and hence}\qquad
\mu_c=0.
\]

It follows from \eqref{eq:isolated_costate_jump} that $\mathbf{p(t_c^-)}=\mathbf{p(t_c^+)}$. The minimum principle then gives
\[
\mathbf{u}^{-}=\mathbf{u}^{+},
\]
so an isolated contact cannot change the leader heading.

Applying the same argument at every isolated contact, and using Lemma~1 on each intervening unconstrained arc, shows that the leader heading is constant over the entire trajectory. The trajectory is therefore the direct straight-line segment from $\mathbf{x}_{L0}$ to $\mathbf{x}_g$, again contradicting its assumed infeasibility.

Hence, the active set $\mathcal{A}$ must contain an interval of non-zero duration.
\end{proof}

\begin{lemma}[Single Boundary Arc]
The active set is a single interval; that is, there exists $t_e<t_x$ such that
\begin{equation}
\mathcal{A}=[t_e,t_x].
\end{equation}
\end{lemma}

\begin{proof}
By Lemma~2, the active set contains an interval of positive duration. Suppose, by contradiction, that $\mathcal{A}$ is not connected. Consider two consecutive connected components of $\mathcal{A}$, and let $t_1$ be the terminal time of the first component and $t_2 > t_1$ be the initial time of the second. Then 
\[
\|\mathbf{z}(t_1)\|
=
\|\mathbf{z}(t_2)\|
=
R,
\]
while
\[
\|\mathbf{z}(t)\|>R,
\qquad t\in(t_1,t_2),
\]
where $\mathbf{z} = \mathbf{x}_L-\mathbf{x}_P.$

The interval $(t_1,t_2)$ is unconstrained. By Lemma~1, the leader heading is constant on this interval. Since the pursuer also moves with constant heading $\mathbf{u}_\theta$, the relative velocity is constant, and hence $\mathbf{z}(t)$ is affine. Therefore,
\begin{equation}
\mathbf{z}(t)
=
(1-\alpha)\mathbf{z}(t_1)
+
\alpha\mathbf{z}(t_2),
\qquad
\alpha
=
\frac{t-t_1}{t_2-t_1}
\in(0,1).
\label{eq:relative_chord}
\end{equation}

If $\mathbf{z}(t_1)\neq\mathbf{z}(t_2)$, then using $\|\mathbf{z}(t_1)\|=\|\mathbf{z}(t_2)\|=R$ gives
\begin{equation}
\|\mathbf{z}(t)\|^2
=
R^2
-
\alpha(1-\alpha)
\|\mathbf{z}(t_2)-\mathbf{z}(t_1)\|^2
<
R^2,
\end{equation}
for every $t\in(t_1,t_2)$. This violates the safety constraint.

If instead $\mathbf{z}(t_1)=\mathbf{z}(t_2)$, then \eqref{eq:relative_chord} gives
\[
\mathbf{z}(t)=\mathbf{z}(t_1),
\qquad t\in[t_1,t_2],
\]
so that $\|\mathbf{z}(t)\|=R$ throughout the interval. This contradicts the assumption that $(t_1,t_2)$ is unconstrained.

Thus, two distinct active components cannot be separated by an unconstrained interval. Hence $\mathcal{A}$ is connected. Since $\mathcal{A}$ is closed and contains an interval of positive duration, there exist $t_e<t_x$ such that $\mathcal{A}=[t_e,t_x]$. 
\end{proof}

\subsection{Boundary Motion}
On the active interval $[t_e,t_x]$, let $\phi(t)$ denote the angular position of the leader relative to the pursuer. Then
\begin{equation}
\mathbf{x}_L(t)
=
\mathbf{x}_P(t)+R\mathbf{u}_{\phi(t)},
\end{equation}
where
\[
\mathbf{u}_{\phi}
=
\begin{bmatrix}
\cos\phi\\
\sin\phi
\end{bmatrix},
\qquad
\boldsymbol{\tau}_{\phi}
\triangleq
\begin{bmatrix}
-\sin\phi\\
\cos\phi
\end{bmatrix}.
\]
Differentiating the active-constraint relation gives
\begin{equation}
\dot{\mathbf{x}}_L
=
v\mathbf{u}_{\theta}
+
R\dot{\phi}\,\boldsymbol{\tau}_{\phi}.
\end{equation}
Using $\|\dot{\mathbf{x}}_L\|=v$ yields
\begin{equation}
R\dot{\phi}
\left[
R\dot{\phi}-2v\sin(\phi-\theta)
\right]
=0.
\end{equation}
Hence, the boundary motion admits two branches:
\begin{equation}
\dot{\phi}=0,
\qquad\text{or}\qquad
\dot{\phi}
=
\frac{2v}{R}\sin(\phi-\theta).
\label{eq:boundary_branches}
\end{equation}

The first branch corresponds to co-moving motion with the pursuer, for which the relative position remains constant. The second branch corresponds to the motion along the safety boundary with a varying relative angle $\phi$; we refer to this as the traversing branch. Using
\[
\mathbf{u}_{2\phi-\theta}-\mathbf{u}_{\theta}
=
2\sin(\phi-\theta)\boldsymbol{\tau}_{\phi},
\]
together with \eqref{eq:boundary_branches}, gives
\begin{equation}
\dot{\mathbf{x}}_L
=
v\mathbf{u}_{2\phi-\theta}.
\end{equation}
Therefore, on the traversing branch,
\begin{equation}
\psi_L
=
2\phi-\theta .
\end{equation}

\begin{lemma}[Boundary-Arc Motion]
Along the minimum-time boundary arc, the co-moving branch $\dot{\phi}=0$ cannot occur over an interval of positive duration. Consequently,
\begin{equation}
\dot{\phi}
=
\frac{2v}{R}\sin(\phi-\theta)
\end{equation}
everywhere on $[t_e,t_x]$.
\end{lemma}

\begin{proof}
By Lemma~3, $\mathcal{A}=[t_e,t_x]$, and $\|x_L(t_f)-x_P(t_f)\|>R$. Hence, $(t_x,t_f]$ is an unconstrained arc. Let $\mathbf{u}_f$ denote the constant leader direction on this final unconstrained arc. By Lemma~1, the costate is constant on this arc and the minimum principle gives
\begin{equation}
\mathbf{p}
=
-\rho\,\mathbf{u}_f,
\qquad
\rho\triangleq\|\mathbf{p} \|\ >0.
\end{equation}

In the relative coordinates, the terminal condition $\mathbf{x}_L(t_f)=\mathbf{x}_g$ is written as
\begin{equation}
\mathbf{F}(\mathbf{z}(t_f),t_f)
=
\mathbf{z}(t_f)
-\mathbf{x}_g
+\mathbf{x}_{P0}
+vt_f\mathbf{u}_\theta
=
\mathbf{0}.
\end{equation}
Since the terminal time is free, the transversality condition ~\cite{kirk1970optimal} gives
\begin{equation}
H_R(t_f)
+
\mathbf{p}(t_f)^\top
v\mathbf{u}_\theta
=
0.
\label{eq:terminal_transversality}
\end{equation}
Using \eqref{eq:HR} and $\mathbf{u}_{\psi_L}(t_f)=\mathbf{u}_f$, \eqref{eq:terminal_transversality} becomes
\[
1+\mathbf{p}(t_f)^\top v\mathbf{u}_f=0.
\]
Substituting $\mathbf{p}=-\rho\mathbf{u}_f$ yields
\[
1-\rho v=0,
\]
and therefore
\begin{equation}
\rho=\frac{1}{v}.
\end{equation}

Hence, on the final unconstrained arc,
\begin{align}
H_R
&=
1-\rho v
\left(
1-\mathbf{u}_f^\top\mathbf{u}_\theta
\right) \nonumber\\
&=
\mathbf{u}_f^\top\mathbf{u}_\theta
\triangleq c.
\label{eq:relative_H_constant}
\end{align}
Since $\mathbf{u}_f$ and $\mathbf{u}_\theta$ are unit vectors, $c\leq1$. If $c=1$, then $\mathbf{u}_f=\mathbf{u}_\theta$, and therefore
\[
\dot{\mathbf{z}}
=
v(\mathbf{u}_f-\mathbf{u}_\theta)
=
\mathbf{0}
\]
on the final unconstrained arc. Since $\|\mathbf{z}(t_x)\|=R$, this would imply $\|\mathbf{z}(t)\|=R$ for all $t\in[t_x,t_f]$, contradicting the strict terminal safety condition \eqref{eq:endpoint_safety}. Thus,
\begin{equation}
c<1.
\label{eq:c_less_than_one}
\end{equation}

The relative dynamics and the state constraint are autonomous. On the boundary arc, the first-order state constraint is incorporated through the augmented Hamiltonian
\[
\overline{H}_R
=
H_R+\nu(t)\dot{g}.
\]
Since $\dot{g}=0$ along the active boundary, $\overline{H}_R=H_R$. Moreover, Hamiltonian continuity at the free--boundary junctions implies that the value in \eqref{eq:relative_H_constant} extends to the boundary arc. Hence,
\begin{equation}
H_R=c<1,
\qquad
t\in[t_e,t_x].
\end{equation}

Suppose now that the co-moving branch $\dot{\phi}=0$ occurs over an interval of positive duration. Then $\mathbf{u}_{\psi_L}=\mathbf{u}_\theta$, and therefore
\[
H_R
=
1+\mathbf{p}^{\top}
v(\mathbf{u}_\theta-\mathbf{u}_\theta)
=
1,
\]
which contradicts $H_R=c<1$.

Therefore, the co-moving branch cannot occur over an interval of positive duration, and the traversing branch
\[
\dot{\phi}
=
\frac{2v}{R}\sin(\phi-\theta)
\]
holds everywhere on $[t_e,t_x]$.
\end{proof}

\subsection{Boundary Junctions}

\begin{lemma}[Smooth Boundary Junctions]
At each junction $t_j\in\{t_e,t_x\}$ between an unconstrained arc and the boundary arc, the leader velocity is continuous:
\begin{equation}
\dot{\mathbf{x}}_L^{F}(t_j)
=
\dot{\mathbf{x}}_L^{B}(t_j),
\label{eq:velocity_continuity}
\end{equation}
where the superscripts $F$ and $B$ denote the free and boundary sides of the junction, respectively. Consequently,
\begin{equation}
\psi_j
=
2\phi_j-\theta
\quad (\operatorname{mod}2\pi),
\qquad
t_j\in\{t_e,t_x\}.
\label{eq:junction_heading}
\end{equation}
\end{lemma}

\begin{proof}
Let
\[
\phi_j\triangleq\phi(t_j),
\qquad
\mathbf{n}_j=\mathbf{u}_{\phi_j},
\qquad
\boldsymbol{\tau}_j=\boldsymbol{\tau}_{\phi_j},
\]
and let $\mathbf{p}_j^{F}$ and $\mathbf{p}_j^{B}$ denote the corresponding one-sided costates.

At a state-constraint junction, the costate jump is normal to the constraint boundary. Hence, its tangential component is continuous:
\begin{equation}
(\mathbf{p}_j^{F})^\top\boldsymbol{\tau}_j
=
(\mathbf{p}_j^{B})^\top\boldsymbol{\tau}_j.
\label{eq:tangential_costate_continuity}
\end{equation}
Since the state constraint has no explicit time dependence in the relative coordinates, the relative Hamiltonian is also continuous across the junction:
\begin{equation}
H_R^{F}(t_j)
=
H_R^{B}(t_j).
\label{eq:HR_junction_continuity}
\end{equation}

On the unconstrained side, the minimum principle gives
\begin{equation}
\mathbf{p}_j^{F}
=
-\rho_j\mathbf{u}_{\psi_j},
\qquad
\rho_j>0,
\label{eq:free_costate}
\end{equation}
where $\mathbf{u}_{\psi_j}$ is the heading of the adjacent unconstrained arc. Therefore,
\begin{equation}
H_R^{F}
=
1-\rho_jv
\left(
1-\mathbf{u}_{\psi_j}^{\top}\mathbf{u}_{\theta}
\right).
\label{eq:HR_free}
\end{equation}

Let
\[
q_j
\triangleq
R\dot{\phi}(t_j^{B})
\]
denote the signed relative tangential velocity on the boundary side. Since
\[
\dot{\mathbf{z}}(t_j^{B})
=
q_j\boldsymbol{\tau}_j,
\]
the corresponding leader velocity is
\begin{equation}
\dot{\mathbf{x}}_L^{B}(t_j)
=
v\mathbf{u}_{\theta}
+
q_j\boldsymbol{\tau}_j.
\label{eq:boundary_velocity}
\end{equation}
Moreover,
\begin{equation}
H_R^{B}
=
1+(\mathbf{p}_j^{B})^\top q_j\boldsymbol{\tau}_j
=
1+q_j(\mathbf{p}_j^{F})^\top\boldsymbol{\tau}_j
=
1-\rho_jq_j
\mathbf{u}_{\psi_j}^{\top}\boldsymbol{\tau}_j .
\label{eq:HR_boundary}
\end{equation}
where \eqref{eq:tangential_costate_continuity} and
\eqref{eq:free_costate} have been used.

Combining
\eqref{eq:HR_junction_continuity},
\eqref{eq:HR_free}, and
\eqref{eq:HR_boundary} gives
\begin{equation}
\mathbf{u}_{\psi_j}^{\top}
\left(
v\mathbf{u}_{\theta}
+
q_j\boldsymbol{\tau}_j
\right)
=
v.
\label{eq:junction_dot_product}
\end{equation}
The vector in parentheses is the boundary-side leader velocity in \eqref{eq:boundary_velocity} and therefore has magnitude $v$. Since $\|\mathbf{u}_{\psi_j}\|=1$, equality in the Cauchy--Schwarz inequality implies
\[
v\mathbf{u}_{\theta}
+
q_j\boldsymbol{\tau}_j
=
v\mathbf{u}_{\psi_j}.
\]
Using \eqref{eq:boundary_velocity}, we obtain
\[
\dot{\mathbf{x}}_L^{B}(t_j)
=
v\mathbf{u}_{\psi_j}
=
\dot{\mathbf{x}}_L^{F}(t_j),
\]
which proves \eqref{eq:velocity_continuity}.

By Lemma~4, the boundary arc follows the traversing branch, for which
\[
\dot{\mathbf{x}}_L^{B}
=
v\mathbf{u}_{2\phi-\theta}.
\]
Evaluating this relation at the junction and using velocity continuity gives
\[
\mathbf{u}_{\psi_j}
=
\mathbf{u}_{2\phi_j-\theta}.
\]
Therefore,
\[
\psi_j
=
2\phi_j-\theta
\quad (\operatorname{mod}2\pi),
\]
which proves \eqref{eq:junction_heading}. 
\end{proof}

The preceding results establish the structure of the minimum-time
trajectory.

\begin{theorem}[Three-Phase Minimum-Time Trajectory]
The minimum-time leader trajectory consists of exactly three phases: an initial straight segment, a single boundary arc, and a final straight segment to the goal.
\end{theorem}

\begin{proof}
Lemma~1 shows that every unconstrained arc is a straight-line segment. Lemma~2 guarantees the existence of a boundary arc of non-zero duration, while Lemma~3 shows that the active set is the single interval $[t_e,t_x]$. Lemma~4 establishes the traversing boundary motion, and Lemma~5 gives velocity continuity at the entry and exit junctions. Together, these results yield the stated three-phase structure.
\end{proof}

\section{Trajectory Construction and Value Function}

The results of Section~III reduce the original problem to constructing the three-phase trajectory for each feasible $\theta$, followed by a one-dimensional optimization over the engagement angle.

\subsection{Boundary Entry}
Let
\[
\mathbf{d}_0
\triangleq
\mathbf{x}_{L0}-\mathbf{x}_{P0},
\qquad
\phi_e\triangleq\phi(t_e).
\]
By Lemma~5, the initial straight segment joins the boundary arc smoothly. Hence, its heading satisfies
\begin{equation}
\psi_e
=
2\phi_e-\theta
\quad (\operatorname{mod}2\pi).
\label{eq:entry_heading}
\end{equation}
Therefore, the leader trajectory before entry is
\begin{equation}
\mathbf{x}_L(t)
=
\mathbf{x}_{L0}
+
vt\,\mathbf{u}_{2\phi_e-\theta},
\qquad
0\leq t\leq t_e.
\label{eq:entry_free_arc}
\end{equation}

At $t=t_e$, the leader lies on the moving safety boundary, so
\begin{equation}
\mathbf{x}_{L0}
+
vt_e\mathbf{u}_{2\phi_e-\theta}
=
\mathbf{x}_{P0}
+
vt_e\mathbf{u}_{\theta}
+
R\mathbf{u}_{\phi_e}.
\label{eq:entry_position}
\end{equation}
Using
\begin{equation}
\mathbf{u}_{2\phi_e-\theta}
-
\mathbf{u}_{\theta}
=
2\sin(\phi_e-\theta)\boldsymbol{\tau}_{\phi_e},
\label{eq:entry_identity}
\end{equation}
equation~\eqref{eq:entry_position} becomes
\begin{equation}
\mathbf{d}_0
+
2vt_e\sin(\phi_e-\theta)
\boldsymbol{\tau}_{\phi_e}
=
R\mathbf{u}_{\phi_e}.
\label{eq:entry_relative}
\end{equation}

Taking the inner product of \eqref{eq:entry_relative} with
$\mathbf{u}_{\phi_e}$ yields
\begin{equation}
\mathbf{d}_0^\top\mathbf{u}_{\phi_e}=R.
\label{eq:entry_angle_condition}
\end{equation}
Thus, the entry angle is determined solely by the initial relative position. Let
\begin{equation}
\rho_0=\|\mathbf{d}_0\|,
\qquad
\beta_0=\operatorname{atan2}(d_{0,y},d_{0,x}).
\end{equation}
Since the leader initially lies outside the safety boundary,
$\rho_0>R$, and \eqref{eq:entry_angle_condition} gives the two candidate entry angles
\begin{equation}
\phi_e^{\pm}
=
\beta_0
\pm
\cos^{-1}\!\left(\frac{R}{\rho_0}\right)
\quad (\operatorname{mod}2\pi).
\label{eq:entry_angle_candidates}
\end{equation}

Taking instead the inner product of \eqref{eq:entry_relative} with $\boldsymbol{\tau}_{\phi_e}$ gives
\begin{equation}
\mathbf{d}_0^\top\boldsymbol{\tau}_{\phi_e}
+
2vt_e\sin(\phi_e-\theta)
=
0,
\end{equation}
and hence
\begin{equation}
{
t_e
=
-\frac{
\mathbf{d}_0^\top\boldsymbol{\tau}_{\phi_e}
}{
2v\sin(\phi_e-\theta)
}.
}
\label{eq:entry_time}
\end{equation}
Only candidates satisfying $t_e>0$ are admissible.
For an admissible candidate, \eqref{eq:entry_relative} also gives,
for $0\leq t\leq t_e$,
\begin{equation}
\mathbf{x}_L(t)-\mathbf{x}_P(t)
=
R\mathbf{u}_{\phi_e}
+
2v(t-t_e)\sin(\phi_e-\theta)
\boldsymbol{\tau}_{\phi_e}.
\end{equation}
Consequently,
\begin{equation}
\|\mathbf{x}_L(t)-\mathbf{x}_P(t)\|^2
=
R^2
+
4v^2(t-t_e)^2
\sin^2(\phi_e-\theta),
\end{equation}
so that
\begin{equation}
\|\mathbf{x}_L(t)-\mathbf{x}_P(t)\|>R,
\qquad
0\leq t<t_e.
\end{equation}
The corresponding entry point is therefore
\begin{equation}
\mathbf{x}_e
=
\mathbf{x}_{P0}
+
vt_e\mathbf{u}_{\theta}
+
R\mathbf{u}_{\phi_e}.
\label{eq:entry_point}
\end{equation}

\subsection{Boundary Propagation}
Once the leader reaches the boundary at $(t_e,\phi_e)$, Lemma~4 gives the boundary-angle dynamics
\begin{equation}
\dot{\phi}
=
\frac{2v}{R}\sin(\phi-\theta).
\label{eq:boundary_angle_dynamics}
\end{equation}
Integrating from $(t_e,\phi_e)$ to
$(t,\phi)$ gives
\begin{equation}
\int_{\phi_e}^{\phi}
\frac{d\varphi}{\sin(\varphi-\theta)}
=
\frac{2v}{R}(t-t_e).
\label{eq:boundary_angle_integral}
\end{equation}
We obtain
\begin{equation}
t(\phi)
=
t_e
+
\frac{R}{2v}
\ln
\left|
\frac{
\tan\!\left(\dfrac{\phi-\theta}{2}\right)
}{
\tan\!\left(\dfrac{\phi_e-\theta}{2}\right)
}
\right|.
\label{eq:boundary_time_angle}
\end{equation}
Accordingly, the leader position at the time $t(\phi)$ corresponding to the boundary angle $\phi$ is
\begin{equation}
    \mathbf{x}_L\bigl(t(\phi)\bigr)
    =
    \mathbf{x}_{P0}
    + v t(\phi)\mathbf{u}_{\theta}
    + R\mathbf{u}_{\phi}.
    \label{eq:boundary_position}
\end{equation}

\subsection{Boundary Exit}
Let $\phi_x\triangleq\phi(t_x)$ denote the boundary angle at exit. By Lemma~5, the final straight segment leaves the boundary smoothly, and hence its heading satisfies
\begin{equation}
\psi_x
=
2\phi_x-\theta
\quad (\operatorname{mod}2\pi).
\label{eq:exit_heading}
\end{equation}
Since this straight segment terminates at the goal,
\begin{equation}
\mathbf{x}_g
=
\mathbf{x}_x
+
v(t_f-t_x)\mathbf{u}_{2\phi_x-\theta},
\label{eq:goal_from_exit}
\end{equation}
where
\begin{equation}
\mathbf{x}_x
=
\mathbf{x}_{P0}
+
v\,t(\phi_x)\mathbf{u}_{\theta}
+
R\mathbf{u}_{\phi_x}.
\label{eq:exit_position}
\end{equation}

Equation~\eqref{eq:goal_from_exit} requires the vector from the exit point to the goal to be parallel to the exit heading. Therefore,
\begin{equation}
\boldsymbol{\tau}_{2\phi_x-\theta}^{\top}
\left(
\mathbf{x}_g-\mathbf{x}_x
\right)
=
0.
\label{eq:exit_collinearity}
\end{equation}
since $\boldsymbol{\tau}_{2\phi_x-\theta}$ denotes the direction perpendicular to $\mathbf{u}_{2\phi_x-\theta}$.

Substituting \eqref{eq:exit_position} gives the scalar exit equation \eqref{eq:exit_collinearity} yields
\begin{equation}
\boldsymbol{\tau}_{2\phi_x-\theta}^{\top}
\left[
\mathbf{x}_g-\mathbf{x}_{P0}
-v\,t(\phi_x)\mathbf{u}_{\theta}
-R\mathbf{u}_{\phi_x}
\right]
=
0.
\label{eq:exit_angle_condition}
\end{equation}
An admissible solution of \eqref{eq:exit_angle_condition} must also satisfy
\begin{equation}
\mathbf{u}_{2\phi_x-\theta}^{\top}
\left[
\mathbf{x}_g-\mathbf{x}_{P0}
-v\,t(\phi_x)\mathbf{u}_{\theta}
-R\mathbf{u}_{\phi_x}
\right]
>0,
\label{eq:exit_forward_condition}
\end{equation}
together with
\begin{equation}
t_x=t(\phi_x)>t_e.
\label{eq:exit_time_condition}
\end{equation}
The duration of the final straight segment is
\begin{equation}
t_f-t_x
=
\frac{\|\mathbf{x}_g-\mathbf{x}_x\|}{v}.
\label{eq:final_segment_time}
\end{equation}
Moreover, the final straight segment remains strictly outside the safety boundary. Indeed, for $t\geq t_x$,
\begin{equation}
    \small
\begin{aligned}
\mathbf{x}_L(t)-\mathbf{x}_P(t)
&=
R\mathbf{u}_{\phi_x}
+
v(t-t_x)
\left(
\mathbf{u}_{2\phi_x-\theta}
-
\mathbf{u}_{\theta}
\right) \\
&=
R\mathbf{u}_{\phi_x}
+
2v(t-t_x)\sin(\phi_x-\theta)
\boldsymbol{\tau}_{\phi_x}.
\end{aligned}
\end{equation}
Hence,
\begin{equation}
\|\mathbf{x}_L(t)-\mathbf{x}_P(t)\|^2
=
R^2
+
4v^2(t-t_x)^2
\sin^2(\phi_x-\theta),
\label{eq:post_exit_safety}
\end{equation}
which is strictly greater than $R^2$ for $t>t_x$ on the traversing branch.

For a fixed $(x_{L0},\theta)$, the entry equation admits at most two 
candidate entry angles, and each may yield one or more admissible exit angles; all feasible entry--exit pairs are retained for the 
value-function construction below.

\subsection{Value Function}
For a given engagement angle $\theta$, let $\mathcal{C}(\mathbf{x}_{L0},\theta)$ denote the set of admissible entry--exit pairs $(\phi_e,\phi_x)$ satisfying the entry condition, the boundary propagation relation, and the exit conditions derived above. For each such pair, the terminal time is
\begin{equation}
t_f
=
t_x
+
\frac{\|\mathbf{x}_g-\mathbf{x}_x\|}{v},
\label{eq:terminal_time}
\end{equation}

The minimum travel time for the immediate-engagement problem is therefore
\begin{equation}
V(\mathbf{x}_{L0},\theta)
=
\min_{(\phi_e,\phi_x)\in
\mathcal{C}(\mathbf{x}_{L0},\theta)}
\left[
t_x+
\frac{\|\mathbf{x}_g-\mathbf{x}_x\|}{v}
\right].
\label{eq:value_function}
\end{equation}
The admissible set $\mathcal{C}(\mathbf{x}_{L0},\theta)$ retains only those entry--exit pairs for which $t_e>0$, $t_x>t_e$, the leader satisfies the safety constraint, and the final segment satisfies the forward-direction condition \eqref{eq:exit_forward_condition}.
Thus, for an escort initially located on the sensing boundary, the minimum-time leader trajectory and its travel time are completely determined for each $\theta$  and each leader starting location $\xlo$. This value function will next be used to treat the case of delayed escort engagement.

\section{Delayed Escort Engagement}
We now relax the assumption from Section~\ref{sec:optimal_structure} and  consider the case in which the escort starts outside the pursuer's sensing boundary. For a fixed engagement angle $\theta$, let $T_E(\theta)>0$ denote the corresponding escort engagement time defined in Section~II. It is assumed that the pursuer remains stationary until $T_E(\theta)$.

\subsection{Reachable Set at Engagement}
During the interval $[0,T_E(\theta)]$, the leader may move while satisfying the safety constraint. Let $\psi_L(\cdot)$ range over all admissible leader heading functions. Define its reachable set at the engagement time by
\begin{equation}
\begin{aligned}
\mathcal{R}_L(\theta)
\triangleq
\bigl\{
\mathbf{y}\in\mathbb{R}^2 \; \big| \;
&
\mathbf{x}_L(0)=\mathbf{x}_{L0},~~
\mathbf{x}_L(T_E(\theta))=\mathbf{y},\\
&
\dot{\mathbf{x}}_L=v\mathbf{u}_{\psi_L},~~
\|\mathbf{x}_L(t)-\mathbf{x}_{P0}\|
\geq R,\;\\
&
0\leq t\leq T_E(\theta)
\bigr\}.
\end{aligned}
\label{eq:leader_reachable_set}
\end{equation}

Thus, at the engagement instant, the leader may occupy any state $\mathbf{y}\in\mathcal{R}_L(\theta)$.

\subsection{Post-Engagement Reduction}
Let $ \tau=t-T_E(\theta) $
denote time measured from the engagement instant. For $\tau\geq0$, the
pursuer motion is
\begin{equation}
\mathbf{x}_P(\tau)
=
\mathbf{x}_{P0}
+
v\tau\mathbf{u}_{\theta}.
\label{eq:post_engagement_pursuer}
\end{equation}
Hence, for any $\mathbf{y}\in\mathcal{R}_L(\theta)$, the continuation problem starting from $ \mathbf{x}_L(0)=\mathbf{y}$
is exactly the immediate-engagement problem considered in Sections~III--IV.
Let
\begin{equation}
\mathcal{R}_L^{3P}(\theta)
\triangleq
\mathcal{R}_L(\theta)
\setminus
\Omega_{\mathrm{straight}}.
\end{equation}
For every
$\mathbf{y}\in\mathcal{R}_L^{3P}(\theta)$, Theorem~1 applies, and the minimum remaining travel time is
$V(\mathbf{y},\theta)$.

\begin{proposition}[Delayed-Engagement Decomposition]
For a fixed engagement angle $\theta$, the optimal leader state at the engagement instant satisfies
\begin{equation}
\mathbf{y}^{\star}(\theta)
\in
\operatorname*{arg\,min}_{
\mathbf{y}\in\mathcal{R}_L^{3P}(\theta)}
V(\mathbf{y},\theta).
\label{eq:optimal_reachable_state}
\end{equation}
The corresponding minimum total travel time is
\begin{equation}
{
J(\mathbf{x}_{L0},\theta)
=
T_E(\theta)
+
V(\mathbf{y}^{\star}(\theta),\theta).
}
\label{eq:delayed_value}
\end{equation}
\end{proposition}

\begin{proof}
The pre-engagement duration $T_E(\theta)$ is fixed for a given $\theta$. Therefore, minimizing the total arrival time is equivalent to selecting, among the leader states reachable at engagement, the state that minimizes the remaining post-engagement travel time. For every $\mathbf{y}\in\mathcal{R}_L^{3P}(\theta)$, this remaining time is $V(\mathbf{y},\theta)$ by the construction in Section~IV, which gives \eqref{eq:optimal_reachable_state} and \eqref{eq:delayed_value}.
\end{proof}

In the numerical study, $y^\star(\theta)$ is obtained by minimizing $V(y,\theta)$ over $\mathcal{R}^{3P}_L(\theta)$.
Finally, the optimal engagement angle satisfies
\begin{equation}
\theta^\star
\in
\operatorname*{arg\,min}_{\theta\in\Theta_{\mathrm{feas}}}
J(\mathbf{x}_{L0},\theta).
\label{eq:optimal_engagement_angle}
\end{equation}
The scalar optimization in \eqref{eq:optimal_engagement_angle} is then evaluated over the feasible engagement-angle set \(\Theta_{\mathrm{feas}}\).

\section{Numerical Results}
We illustrate the trajectory structure and the effect of escort engagement using two representative configurations. In the numerical examples, all agents have unit speed, the pursuer starts at \(x_{P0}=(0,0)\), the sensing radius is \(r=1.8\), and the safety radius is \(R=r+\epsilon=2\). For each configuration, we first show a feasible trajectory for a prescribed engagement angle and then optimize the engagement angle over the feasible set. The corresponding minimum-time trajectory without escort assistance is included for comparison.
\begin{figure*}[t]
    \centering
    \setlength{\tabcolsep}{5pt}

    \begin{tabular}{ccc}

        \begin{minipage}[c]{0.25\textwidth}
            \centering
            \includegraphics[width=\linewidth]{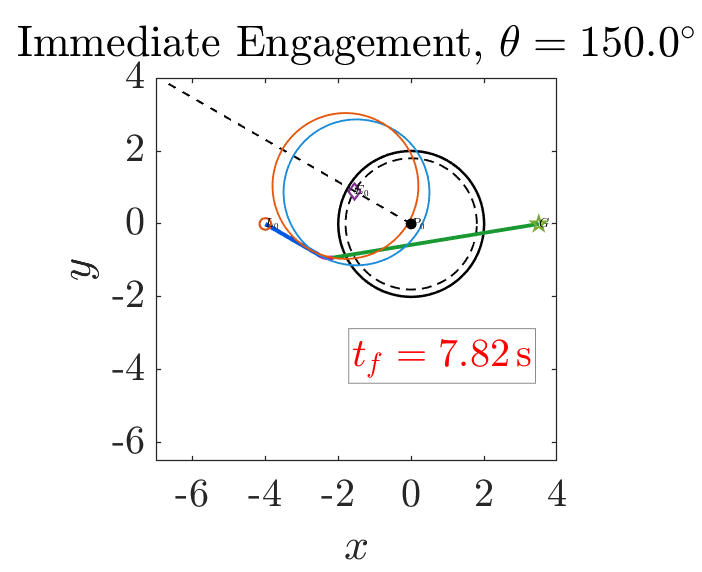}\\[-1mm]
            {\small (a) Immediate, prescribed angle}
        \end{minipage}
        &
        \begin{minipage}[c]{0.25\textwidth}
            \centering
            \includegraphics[width=\linewidth]{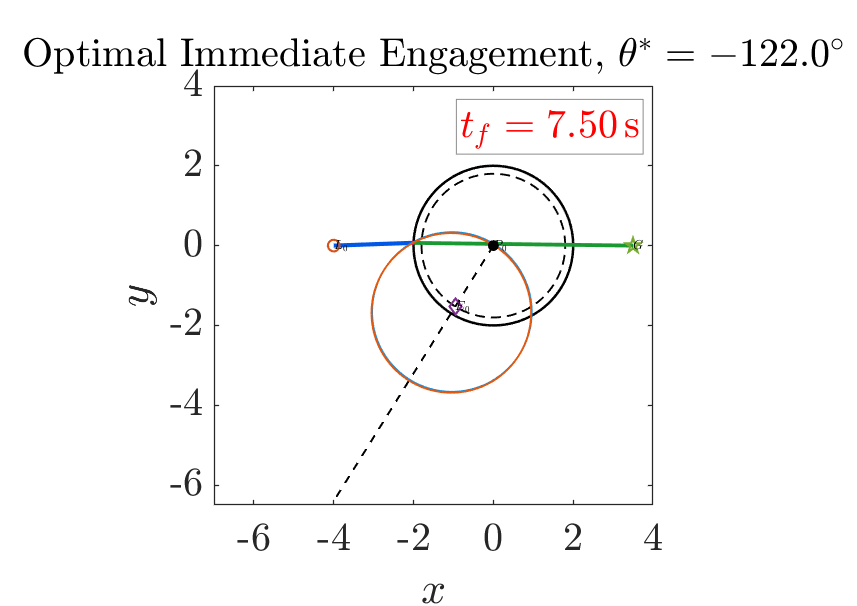}\\[-1mm]
            {\small (b) Immediate, optimal angle}
        \end{minipage}
        &
        \begin{minipage}[c]{0.25\textwidth}
            \centering
            \includegraphics[width=\linewidth]{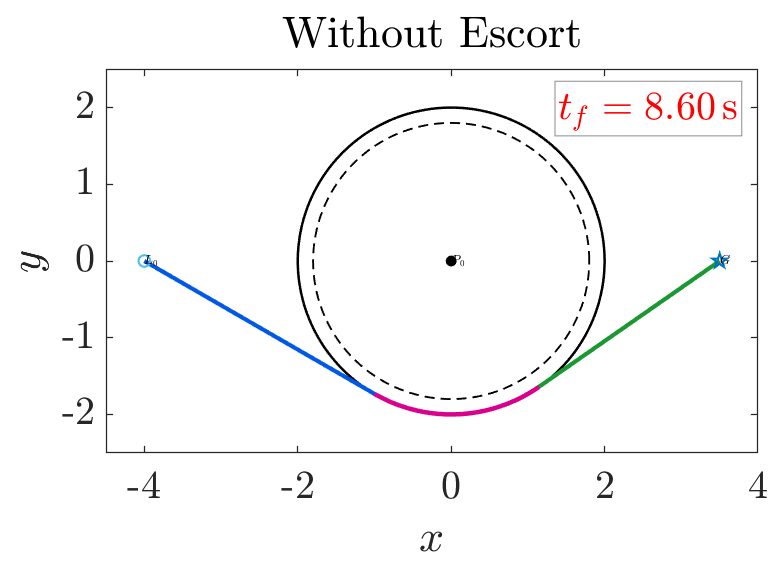}\\[-1mm]
            {\small (e) No escort}
        \end{minipage}
        \\[3mm]

        \begin{minipage}[c]{0.25\textwidth}
            \centering
            \includegraphics[width=\linewidth]{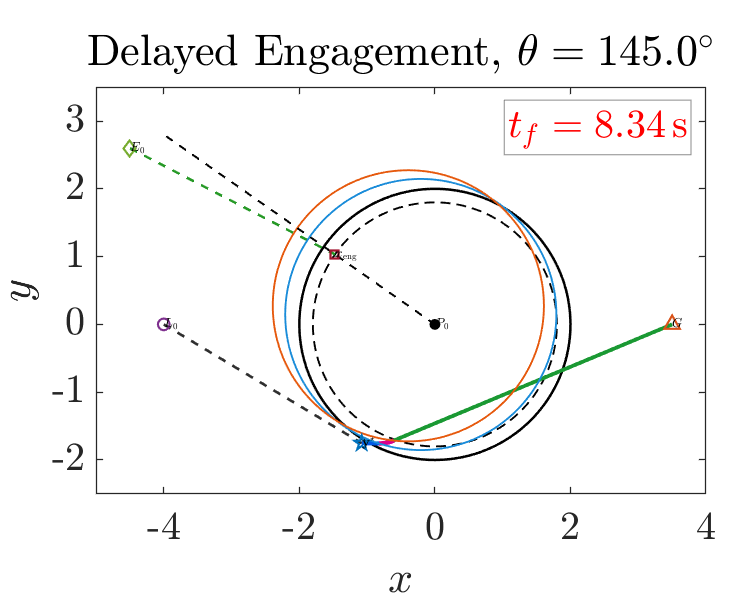}\\[-1mm]
            {\small (c) Delayed, prescribed angle}
        \end{minipage}
        &
        \begin{minipage}[c]{0.25\textwidth}
            \centering
            \includegraphics[width=\linewidth]{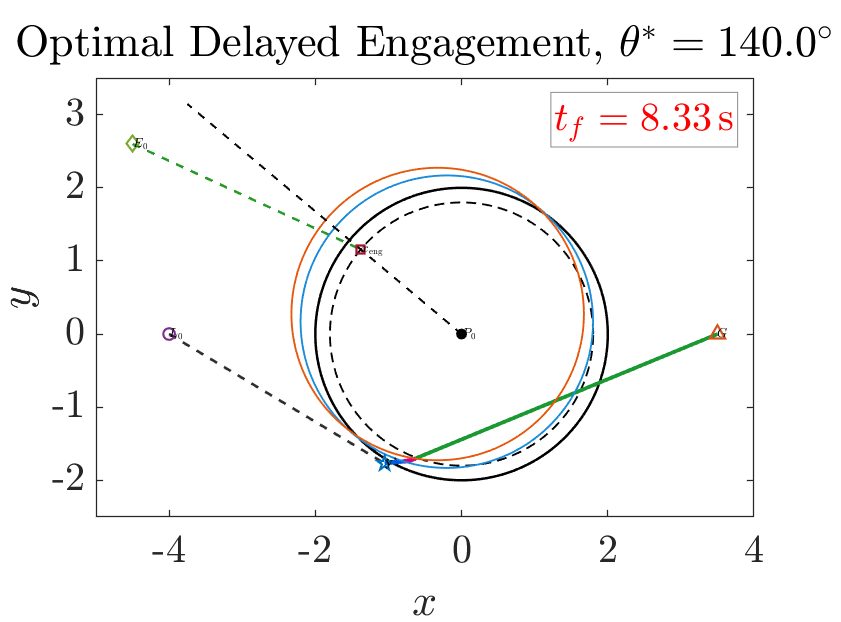}\\[-1mm]
            {\small (d) Delayed, optimal angle}
        \end{minipage}
        &
        \begin{minipage}[c]{0.28\textwidth}
    \small
    \centering
    \textbf{Trajectory notation}\\[2mm]

    \begin{tabular}{cl}

        \rule[0.5ex]{14mm}{0.8pt}
        & Initial safety boundary \\[1mm]

        \raisebox{0.5ex}{\makebox[14mm]{\dotfill}}
        & Sensing boundary \\[1mm]

        \textcolor{blue}{\rule[0.5ex]{14mm}{0.8pt}}
        & Safety boundary at entry \\[1mm]

        \textcolor{orange}{\rule[0.5ex]{14mm}{0.8pt}}
        & Safety boundary at exit \\[1mm]

        \textcolor{blue}{\rule[0.5ex]{14mm}{1.2pt}}
        & Phase 1: initial straight segment \\[1mm]

        \textcolor{magenta}{\rule[0.5ex]{14mm}{1.2pt}}
        & Phase 2: moving-boundary segment \\[1mm]

        \textcolor{green!60!black}{\rule[0.5ex]{14mm}{1.2pt}}
        & Phase 3: final straight segment \\[1mm]

        {\Large \textcolor{blue}{$\ast$}}
        & Leader state $\mathbf{y}^{\star}$ at engagement

    \end{tabular}
\end{minipage}

    \end{tabular}

    \caption{Minimum-time trajectories for immediate and delayed escort
    engagement. Panels (a) and (b) show prescribed-angle and optimal-angle
    trajectories, respectively, for immediate engagement. Panels (c) and
    (d) show the corresponding prescribed-angle and optimal-angle trajectories
    for delayed engagement. Panel (e) shows the corresponding minimum-time
    trajectory without escort assistance.}

    \label{fig:simulation_trajectories}
\end{figure*}

\subsection{Immediate Engagement}
For $T_E=0$, Fig.~\ref{fig:simulation_trajectories}(a) shows a feasible
three-phase trajectory for $\theta=150^\circ$. Optimizing the engagement
angle gives Fig.~\ref{fig:simulation_trajectories}(b), with
$\theta^\star=-122^\circ$ and travel time $t_f^\star=7.50\,\mathrm{s}$.
For the same leader initial condition, the no-escort trajectory in
Fig.~\ref{fig:simulation_trajectories}(e) has travel time
$t_f^{\rm NE}=8.60\,\mathrm{s}$. Thus, escort assistance reduces the travel time by
$ 1.10\,\mathrm{s}$.

\subsection{Delayed Engagement}
For delayed engagement, the escort starts outside the sensing boundary
and reaches the engagement point at $T_E>0$.
Figure~\ref{fig:simulation_trajectories}(c) shows a feasible trajectory
for $\theta=145^\circ$. Optimizing the engagement angle gives
Fig.~\ref{fig:simulation_trajectories}(d), with
$\theta^\star=140^\circ$ and total mission time
$t_f^\star=8.33\,\mathrm{s}$.
 Hence, compared with the no-escort case, delayed escort assistance reduces the
travel time by $0.27\,\mathrm{s}$.

The numerical examples illustrate the trajectory structure derived in Sections~III--IV and the benefit of escort assistance.

\section{Conclusion}
This paper developed a minimum-time framework for escort-assisted evasion from a sensing-limited pursuer. We established the optimal straight–boundary–straight trajectory for immediate engagement and extended the solution to delayed engagement using reachable-set analysis.

Several extensions remain for future work. First, the equal-speed assumption will be relaxed to allow different leader, escort, and pursuer speeds, which may lead to qualitatively different boundary dynamics and trajectory structures. Second, one may consider more general pursuer response models in which the post-engagement motion is not restricted to a prescribed constant heading, including feedback pursuit strategies. Another direction is to incorporate vehicle dynamics and control constraints, such as bounded turn rate and acceleration, in place of the first-order constant-speed model. Finally, extending the framework to multiple escorts and pursuers will enable the study of cooperative engagement and trajectory planning in more complex adversarial environments.

\appendix

\section*{Normality of the Extremals}
\label{app:normality}
We establish the normality assumption used in Section~III. Retain the general PMP multiplier $\lambda_0\geq0$ and suppose, by contradiction, that $\lambda_0=0$. On an unconstrained arc, \eqref{eq:costate_dot} gives $\dot{\mathbf p}=0$. Since $\lambda_0$ does not affect the minimization with respect to the control, whenever $\mathbf p\neq0$ the minimum principle gives $ \mathbf u_{\psi_L}^{\star} =-\frac{\mathbf p}{\|\mathbf p\|}, $ as in \eqref{unconstrained_u} 

Using the terminal transversality condition \eqref{eq:terminal_transversality} with the general Hamiltonian \eqref{eq:HR} gives $ \lambda_0 +\mathbf p(t_f)^\top v\mathbf u_{\psi_L(t_f)}=0. $ Under $\lambda_0=0$, \eqref{unconstrained_u} therefore yields
\[
    -v\|\mathbf p(t_f)\|=0,
\]
so that $\mathbf p(t_f)=0$. Since the terminal safety constraint is strictly inactive, the terminal arc is unconstrained, and \eqref{eq:costate_dot} implies that $\mathbf p$ vanishes throughout that arc.

For isolated contacts, the junction analysis in the proof of Lemma~2 shows that the contact multiplier vanishes and the costate is continuous. For a boundary interval, the free--boundary junction conditions used in Lemma~5 apply at entry and exit. Hence an abnormal\ extremal cannot retain a nontrivial multiplier through the finite sequence of free and active arcs, contradicting the PMP nontriviality condition. Therefore $\lambda_0>0$, and by homogeneity we normalize
\begin{equation}
    \lambda_0=1.
\end{equation}
Consequently, $\mathbf p\neq0$ on every unconstrained arc.

\bibliographystyle{IEEEtran}
\bibliography{references}
\end{document}